\documentclass[twocolumn,prl,showpacs,nobibnotes,preprintnumbers]{revtex4}

\usepackage{amsmath}
\usepackage{amssymb}
\usepackage{amsfonts}
\usepackage{graphicx}

\newcommand{\eg}{e.\,g.\ }
\newcommand{\ie}{i.\,e.\ }

\begin{document}

\title{Versatile approach to access the low temperature
  thermodynamics of lattice polymers and proteins}

\author{Thomas W\"ust}
\email{twuest@physast.uga.edu}
\author{David P.\ Landau}
\affiliation{Center for Simulational Physics, The University of
  Georgia, Athens, GA 30602, USA}


\begin{abstract}
We show that Wang-Landau sampling, combined with suitable Monte Carlo
trial moves, provides a powerful method for both the ground state
search and the determination of the density of states for the
hydrophobic-polar (HP) protein model and the interacting self-avoiding
walk (ISAW) model for homopolymers. We obtained accurate estimates of
thermodynamic quantities for HP sequences with $>100$ monomers and for
ISAWs up to $>500$ monomers. Our procedure possesses an intrinsic
simplicity and overcomes the limitations inherent in more tailored
approaches making it interesting for a broad range of protein and
polymer models.
\end{abstract}


\maketitle

\begin{figure}
\includegraphics[width=0.85\columnwidth]{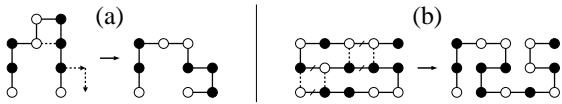}
\caption{\label{figure1}Typical example of pull (a) and
  bond-rebridging (b) move in 2D. For details, see
  \cite{pull,cutjoin}.}
\end{figure}

Coarse-grained polymer and protein models play an important role in
understanding physical phenomena such as \eg protein folding or the
phase behavior of flexible macromolecules, and Monte Carlo simulation
methods have become an indispensable tool for the study of such models
\cite{models}. One of the most prominent examples is the
hydrophobic-polar (HP) lattice model \cite{hpmodel}, where the protein
is represented as a self-avoiding chain of beads (the amino acid
residues) on a lattice. The amino acids are divided into two classes -
hydrophobic (H) and polar (P) - and an attractive interaction
$\epsilon$ acts between non-bonded neighboring H residues mimicking
the hydrophobic force ($\epsilon_{HH}=-1, \epsilon_{HP,PP}=0$). The
special case of a chain consisting entirely of H residues
(homopolymer), the interacting self-avoiding walk (ISAW), is
an important model for studying the statistical physics of polymers
\cite{sokal:95,polyreview}.\\
Despite their formal simplicity and minimalistic framework, lattice
models represent a challenging testing ground for computational methods
because of their complex energy landscapes, conformational constraints
and dense packings. The HP model has become a standard for assessing
the efficiency of folding algorithms, and numerous - some very tailored
- conformational ground state search strategies have been proposed,
see \eg \cite{perm,pull,cpsp,fress} and references therein.\\
More revealing than algorithms that merely search for low energy
states, however, are methods which target the sampling of the entire
conformation and energy space. They can provide an estimate of the
density of states (DOS) $g(E)$ of energy $E$ which, in turn, gives access to thermodynamic
properties (\eg internal energy, specific heat, entropy or free
energy) of a system at any temperature \cite{landau:05}. Only a few
attempts have been undertaken to this end for the HP model, the most
notable approaches being multi-self overlap ensemble Monte Carlo
(MSOE) \cite{msoe}, multicanonical chain growth (MCCG) \cite{mccg},
and equi-energy sampling (EES) \cite{ees}. Although inventive and powerful,
these methods also suffer from severe limitations: Large memory
needs for keeping track of all sampled conformations (construction of
microcanonical ensembles) (EES); (quasi-) statics, \ie one bead of the
chain is permanently fixed in space (MCCG); or the necessity to treat
an expanded ensemble resulting in a large amount of computer time
spent in sampling non-physical space (MSOE). Such restrictions can
become increasingly important for more complex biological
setups such as multi-chain systems or protein folding in heterogeneous
environments (\eg membranes) \cite{HPexamples}.\\
In this Letter we show that a generic algorithm - Wang-Landau sampling
\cite{wls} - together with appropriate Monte Carlo trial moves,
provides a powerful, yet flexible methodology for the simulation of
HP-like lattice proteins and homopolymers that does not suffer from
any of the above limitations.\\
The key to our approach is the combination of two ``non-traditional''
Monte Carlo trial moves, which complement one another extremely well,
namely pull moves \cite{pull} and bond-rebridging moves \cite{cutjoin}, see Fig.~\ref{figure1}.
Originally
proposed for the square and simple cubic lattices only, here
we extended both types of trial moves to any n-dimensional space
($n\ge 2$). (i) Pull moves \cite{pull} allow
for the close-fitting motion of a polymer chain within a confining
environment by ``pulling'' portions of the polymer to unoccupied
neighboring sites. Pull moves are
reversible and fulfill ergodicity; moreover, they provide a good
balance between local and global conformational changes, as well as a
``natural'' dynamics of folding. These features are important to an
algorithm that seeks to sample the entire conformational space such as
Wang-Landau sampling and thus requires an efficient move for the
continuous folding and unfolding of the polymer. (ii) Bond-rebridging
moves \cite{cutjoin}: Trial moves which displace monomers become
ineffective for very compact conformations where few unoccupied
neighboring sites remain available. In contrast, bond-rebridging moves
allow the polymer to change its conformation even at highest densities
by reordering bonds while leaving the positions of monomers
unchanged. Moreover, they facilitate long range topological changes,
\eg entanglement, which otherwise require costly unfolding/folding
processes. This later feature becomes particularly important when the
sampling of the DOS is split up into energy subintervals
as it substantially reduces the risk of ``locking-out'' conformational
space \cite{adaptWL}. During sampling, pull or
bond-rebridging trial moves were selected randomly (usually with a
$1:1$ ratio for each type). Pull moves enabled us to sample the entire
conformational space of long polymer chains which was not feasible with ``traditional
moves'' only \cite{sokal:95}. Furthermore, the combination
of bond-rebridging and pull moves provided a speed-up of a factor 3
for the HP model and a factor of 10 for the ISAW as
comparing with pull moves only.\\
Wang-Landau (WL) sampling is an efficient and robust algorithm for the
computation of the DOS for diverse statistical physical systems, see
\cite{wls,landau:05} for details. To fulfill detailed balance in
conjunction with pull moves, in this study the WL transition
probability from a state $A$ to a state $B$ has been generalized to
\begin{equation}
\label{wl_transition}
P(A\rightarrow B)=
\min\left(1,\frac{g(E_A)}{g(E_B)}\times
\frac{n_{B\rightarrow A}/n_B}{n_{A\rightarrow B}/n_A}\right).
\end{equation}
$n_{A\rightarrow B}$ denotes the number of pull moves from $A$ to $B$
and $n_A$ the total number of possible pull moves from $A$;
$n_{B\rightarrow A}$ and $n_B$ correspondingly (here, $n_{A\rightarrow
  B}=n_{B\rightarrow A}$ because of reversibility). Selecting only
within the list of possible pull moves ($n_A$) also increases the dynamics for
dense conformations as compared to a standard ``trial and
error'' procedure. In order to yield
accurate and reliable DOS estimates over the entire energy range
(including the lowest energies) we used a very stringent parameter set
for all our simulations, \ie final modification factor
$\ln(f_{\text{final}})=10^{-8}$ and flatness criterion $p=0.8$;
statistical errors were always calculated from 15 independent DOS
estimates (by means of a Jackknife analysis).\\
\begin{figure}
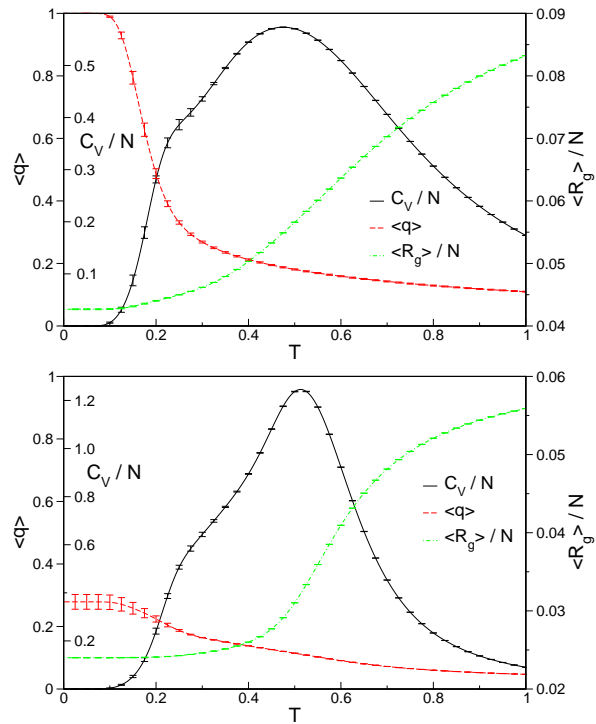

\includegraphics[width=0.90\columnwidth]{./figure2a.eps}\\[0.5ex]
\includegraphics[width=0.90\columnwidth]{./figure2b.eps}
\caption{\label{figure2}(Color online) Specific heat $C_V/N$, mean
  radius of gyration $\langle R_g\rangle/N$ ($N$, chain length), and
  mean Jaccard index $\langle q\rangle$ as a function of temperature
  ($T$) for HP sequence 2D100b (\emph{top}) and 3D103 (\emph{bottom}),
  respectively.}
\end{figure}
The knowledge of the exact energy range is essential in the WL
algorithm for the examination of the flatness of the histogram. Often,
however, energy boundaries are \emph{a priori} unknown, (hence the
use of ground state search algorithms, \eg for the HP model). To solve
this dilemma, the following procedure proved to be most efficient:
Every time a new energy level $E_{\text{new}}$ is found, it is marked
as ``visited'' and $g(E_{\text{new}})$ is set to
$g_{\text{min}}$, \ie the minimum of $g$ among all previously visited
energy levels. The flatness of the histogram is checked for visited
energy levels only. With this self-adaptive procedure, new regions of
conformational space can be explored while, at the same time, the
current DOS estimate is further refined.\\
First, we applied our procedure to various benchmark HP sequences
found in the literature. Since heteropolymers with $N\le 50$ no longer
represent a significant challenge and our results are in perfect
agreement with previous works, we restrict our presentation to two
longer sequences which turned out to be particularly demanding,
namely, a 100mer in 2D (2D100b) and a 103mer in 3D (3D103); for
definitions of HP sequences, see \eg \cite{fress}. The ground states
of sequence 2D100b are believed to have an energy $E=-50$
\cite{msoe,perm,pull,fress}; however, previous attempts to obtain the
DOS over the entire energy range $[-50,0]$ within a single simulation
have failed \cite{msoe,ees}. In contrast, with our approach we were
able to achieve this with high accuracy.
Fig.~\ref{figure2}~(\emph{top}) shows the resulting specific heat
$C_V(T)/N$, depicting a peak at $T\approx 0.48$ (coil-globule
transition) and a very weak shoulder at $T\approx 0.23$ (folding
transition). Such two-step acquisition of the native state
has been observed in studies of realistic protein models and is not
restricted to lattice models. For sequence 3D103, the lowest energy
found so far was $-57$, achieved only by fragment regrowth Monte
Carlo via energy-guided sequential sampling (FRESS) \cite{fress}. With
our approach, we discovered an even lower state with energy
$-58$. Moreover, we were also able to obtain the DOS in
the energy range $[-57,0]$, within a single simulation, and with
very high accuracy. It was nonetheless not possible to determine the
relative magnitudes of the ground state ($E=-58$) and 1st excited
state ($E=-57$) DOS with high precision.
Fig.~\ref{figure2}~(\emph{bottom}) displays the specific heat for
sequence 3D103, manifesting a peak at $T\approx 0.51$ and a shoulder
at $T\approx 0.27$. We do not observe an additional peak in $C_V$ at
very low temperatures, contrarily to Ref.~\cite{mccg}. However,
since only conformations with energies down to $E=-56$ were found and
the estimated errors near that peak were rather large, we think that
this finding was an artefact of insufficient sampling. Indeed, our
$C_V$ curves indicate that the folding transitions from unstructured
globular conformations to the ground states are rather weak for both
sequences - despite the difficulty in sampling their low energy
regimes.\\
By means of multicanonical sampling given our DOS estimates, we
obtained the radius of gyration $R_g$ \cite{sokal:95} and the Jaccard
index $q=\max\left\{c_{s,g}/(c_{s,g}+c_s+c_g)|E_g=\min\right\}$
which measures the structural similarity
between any conformation $s$ and the ground states $g$ of an HP
sequence \cite{fraser:07}.
$c_{s,g}$ denotes the number of common (native) H-H contacts between
$s$ and $g$, and $c_s$, $c_g$ are the numbers of H-H contacts found
only in $s$ and $g$, respectively (the maximum stems from the possible
degeneracy of ground states). Fig.~\ref{figure2} also shows the
averages $\langle R_g\rangle$ and $\langle q\rangle$ for sequences
2D100b and 3D103 and illustrates the complementary information in
these two quantities. While $\langle R_g\rangle$ indicates the
coil-to-globule collapse, $\langle q\rangle$ identifies the folding
transition to the native state and thus may serve as a suitable
structural order parameter for these kind of systems. In case of
sequence 3D103, the ground state ($E=-58$) was excluded from the
sampling (due to the difficulty in finding this state) which results
in $\langle q\rangle$ saturating at a rather low value ($<0.3$) for
$T\rightarrow 0$. This manifests the still large structural
differences between conformations with $E=-57$ and the ground state.\\
TABLE~\ref{Seqs} compares various methods in finding low energy
conformations and, if available, the DOS for common benchmark HP
sequences. We also included results from methods which were focused on
the low temperature range only, \ie FRESS \cite{fress} and the
variants of PERM (pruned-enriched Rosenbluth method) \cite{perm} and
hence do not provide the entire DOS. Except for the longest sequence
(3D136), we could confirm all minimum energy states found
previously. The superiority of FRESS for this sequence is the result of
various ``efficiency enhancements'' towards low energy states (see
\cite{fress}) which become obviously the more effective the longer the
chain length. However, they do not permit anymore a correct sampling,
let alone an estimation of the DOS.\\
\begin{table}
\caption{\label{Seqs}Energy minima found by several
  methods for benchmark HP sequences in 2D and 3D. The first column
  names the sequence (dimension and length), see \cite{fress}. In case
  of Wang-Landau sampling (WLS), numbers in parentheses denote that
  the DOS has been obtained down to this energy. Horizontal lines mean
  no data available.}
\begin{ruledtabular}
\begin{tabular}{lcccccc}
Seq.   & WLS       & EES & MCCG & MSOE & FRESS\footnotemark[2] & PERM\footnotemark[2]\\\hline
2D100a & -48       & -48 &  --  & -47  &  -48  & -48\\
2D100b & -50       & -49 &  --  & -50\footnotemark[3]  &  -50  & -50\\
3D88   & -72 (-69) &  -- &  --  &  --  &  -72  & -69\\
3D103  & -58\footnotemark[1] (-57) &  -- & -56  &  --  &  -57  & -55\\
3D124  & -75 (-74) &  -- &  --  &  --  &  -75  & -71\\
3D136  & -82 (-81) &  -- &  --  &  --  &  -83  & -80\\
\end{tabular}
\end{ruledtabular}
\footnotetext[1]{See \eg
$d$\-$r_2$\-$u_2$\-$l$\-$d$\-$b$\-$d$\-$r$\-$u$\-$b$\-$d$\-$b$\-$l$\-$f$\-$l$\-$d$\-$r$\-$b$\-$l_2$\-$f_2$\-$d$\-$r_2$\-$d$\-$b$\-$r$\-$u$\-$l$\-$b$\-$r_2$\-$d$\-$r$\-$f$\-$r$\-$u$\-$l_2$\-$d$\-$f$\-$u$\-$r$\-$u$\-$f$\-$d$\-$l$\-$d$\-$f$\-$l$\-$b_2$\-$u$\-$f$\-$u$\-$f_2$\-$r$\-$b_2$\-$u$\-$r$\-$b_2$\-$r$\-$b$\-$l$\-$u$\-$l$\-$d$\-$f$\-$u_2$\-$f$\-$d_3$\-$b$\-$r$\-$b$\-$l$\-$b$\-$u$\-$l_2$\-$b$\-$r$\-$d$\-$f_4$\-$l$\-$f_2$\-$d$\-$b$\-$l$\-$d$
(encoded as sequence left[l], right[r], up[u], down[d], forward[f], backward[b]).}
\footnotetext[2]{Ground state search only (no DOS estimate)}
\footnotetext[3]{DOS not attained.}
\end{table}
As a second test of performance, we applied our method to the
interacting self-avoiding walk (ISAW) representing a homopolymer with
nearest-neighbor attraction ($\epsilon=-1$) on the square (sq, 2D) and
simple cubic (sc, 3D) lattice. Unraveling the ``phase transition''
behavior of flexible macromolecules in the thermodynamic limit
($N\rightarrow\infty$) by means of simple (lattice) models - such as
\eg the ISAW, the bond-fluctuation model or systems in the continuum -
has been a challenge for decades \cite{polyreview,polymer}. Although
the $\theta$ point (coil-globule transition) could be investigated
well for polymer chains with $N\ge 10\,000$ monomers, our understanding
of the ISAW at very low temperatures remains elusive. Due to the very
dense packings resulting for this model, accurate estimates of
thermodynamic quantities below $T_{\theta}$ are difficult to
obtain. In the most recent computational studies, only chains with
$N\le 125$ in 3D (multicanonical chain-growth \cite{polymer_sc}) and
$N\le300$ in 2D (adaptive WL sampling with reptation, but without the
lowest energy states \cite{adaptWL}) could be investigated.\\
\begin{figure}
\includegraphics[width=0.80\columnwidth]{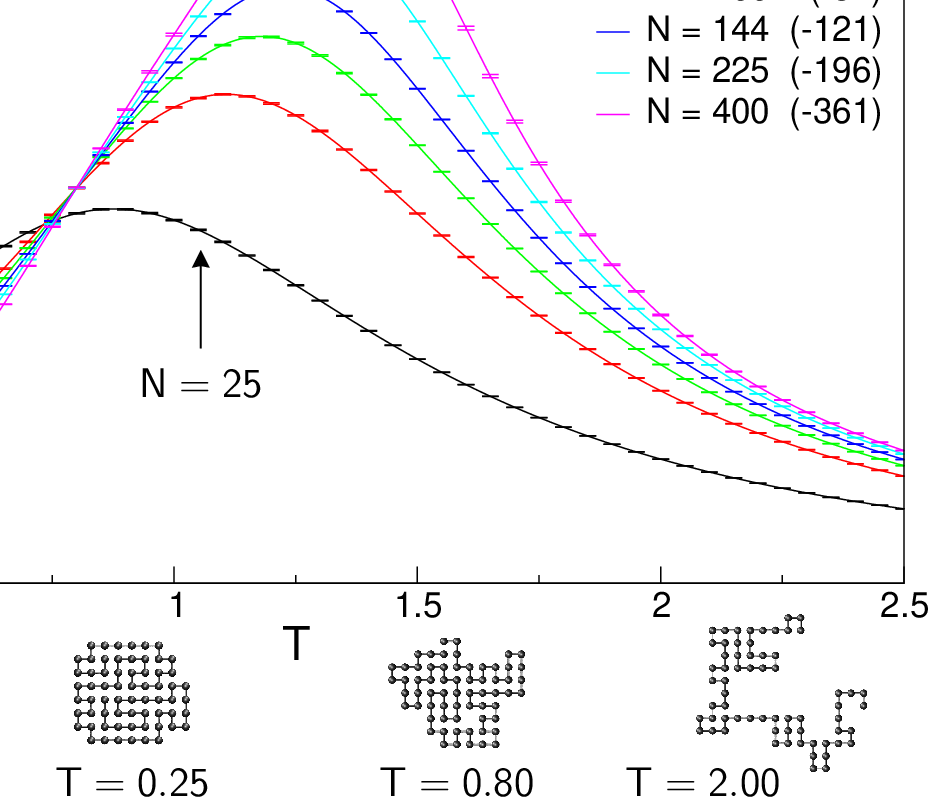}\\[2ex]
\includegraphics[width=0.80\columnwidth]{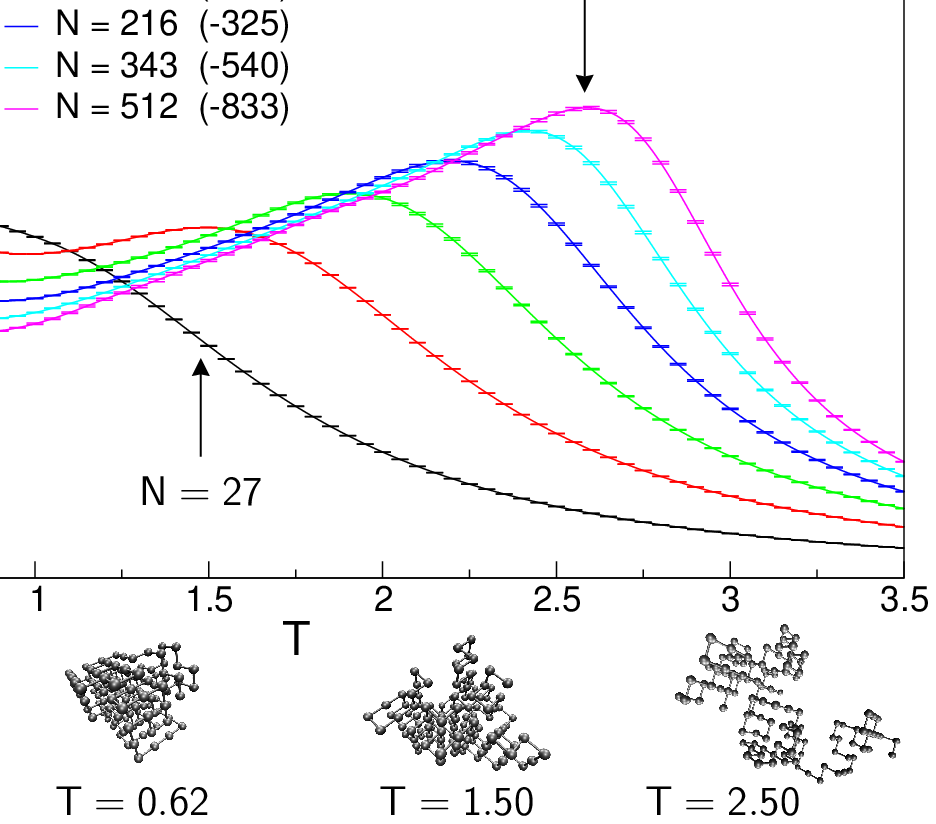}
\caption{\label{figure3}(Color online) Specific heat $C_V/N$ as a
  function of temperature ($T$) for ISAWs of
  various chain lengths $N$ on square (\emph{top}) and simple
  cubic (\emph{bottom}) lattice. Numbers in parentheses denote
  corresponding energy minima. \emph{Bottom rows:} Representative
  structures at specific temperatures for $N=64$ (2D) and $N=125$
  (3D).}
\end{figure}
With our generic approach we were able to obtain accurate DOS
estimates for ISAWs up to chain lengths $N=400$ (2D) and $N=512$ (3D)
over the \emph{entire} energy range (including ground states) and we
could then determine reliable thermodynamic quantities even at lowest
temperatures ($T\rightarrow 0$), see Fig.~\ref{figure3}. The
possibility to compare the specific heat $C_V(T)$ for various system
sizes up to these chain lengths allowed us to draw interesting
conclusions which apply for the ISAW on both the sq and sc lattice: At
high $T$, the collapse transition ($\theta$ point) indicates a clear
phase transition manifested by cooperative structural rearrangements
from the coil to the globular state and
$C_V(T_{\theta})\rightarrow\infty$ for $N\rightarrow\infty$. At very
low $T$, a pronounced peak appears due to various ground state
excitations (here, the ground states form either regular squares or
cubes). These excitations are induced by \emph{local} rearrangements
at the surface and therefore, the magnitude of the peak decreases
systematically with chain length (2D) or becomes constant to within
statistical errors bars (3D). The breaking up of the ground state
structure bears similarity to surface roughening on crystal facets,
\ie the formation of kinks and edges at the surface of a compact core
without vacancies (indeed, bulk vacancies appear at much higher $T$
only). At intermediate temperatures, metastable (and chain length
dependent) phases emerge but they gradually diminish for
$N\rightarrow\infty$. Most notably, the ISAW on the sq/sc lattice does
not undergo a true crystallization transition as observed for other lattice
and off-lattice polymer models \cite{polyreview,polymer}. Once in the
globular phase, the rigidity of the model does not permit a further
cooperative effect (\ie symmetry breaking) which would be necessary
for such a transition. Whereas a variation of chain length
($N\ne\text{``magic'' number}$) has an influence on the magnitude and
the position of the excitation peak at low $T$, the overall
thermodynamic scenario remains the same for sufficiently large
$N$. Note that it was essential to have data for chains that were
longer than other methods could treat in order to ascertain the low
$T$ behavior of the ISAW in the thermodynamic limit.\\
In summary we have shown that Wang-Landau sampling with suitable Monte
Carlo trial moves (pull and bond-rebridging moves combined) offers a
powerful solution for studying the thermodynamics of lattice homo- and
heteropolymers even in the very demanding low temperature ranges of
such models. A major advantage of our method is that it remains rather
simple and flexible beside its proven performance which has not been
achieved earlier, by more elaborate attempts
\cite{msoe,mccg,ees,polymer_sc}. These features make it readily
applicable to the study of complex biological phenomena such as \eg
protein aggregation or protein insertion into a membrane
\cite{HPexamples}. Since both trial moves are usable for lattice and off-lattice models \cite{cutjoincontinuous}, other systems with
conformational constraints should also benefit from our self-adaptive
WL procedure.\\
We thank K.\ Binder and W.\ Paul as well as C.\ Gervais and
D.~T.\ Seaton for fruitful discussions. This work was supported in
part by NSF Grant DMR-0810223.

\end{document}